\documentclass[11pt, authoryear]{elsarticle}              
\journal{arXiv}

\usepackage{amssymb,amsfonts,amsmath}
\usepackage{xcolor}
\usepackage{natbib}
\usepackage{amsmath}
\usepackage[top=1.0in, bottom=1.0in, left=1.0in, right=1.0 in]{geometry}
\usepackage[doublespacing]{setspace}
\usepackage[bottom]{footmisc}
\usepackage{indentfirst}
\usepackage{endnotes}
\usepackage{lastpage}
\usepackage{hyperref,url}                
\usepackage{epsfig,pgf,subfig}       
\usepackage{array}
\usepackage{rotating,longtable,booktabs,colortbl,siunitx}       
\usepackage{mdwlist}
\usepackage{lscape,afterpage}                                   
\usepackage{appendix}

\usepackage{dcolumn}
\newcolumntype{d}[1]{D{.}{.}{#1}}        

\makeatletter    

\def\@author#1{\g@addto@macro\elsauthors{\normalsize%
    \def\baselinestretch{1}%
    \upshape\authorsep#1\unskip\textsuperscript{%
      \ifx\@fnmark\@empty\else\unskip\sep\@fnmark\let\sep=,\fi
      \ifx\@corref\@empty\else\unskip\sep\@corref\let\sep=,\fi
      }%
    \def\authorsep{\unskip,\space}%
    \global\let\@fnmark\@empty
    \global\let\@corref\@empty  
    \global\let\sep\@empty}%
    \@eadauthor={#1}
}

\makeatother     

\begin{document}

\begin{frontmatter}
\title{\textbf{Modeling and Analysis of Discrete Response Data: Applications 
to Public Opinion on Marijuana Legalization in the United States} }

\author[add1]{Mohit Batham}
\ead{mohitbatham15@gmail.com}
\author[add2]{Soudeh Mirghasemi}
\ead{Soudeh.Mirghasemi@hofstra.edu}
\author[add3]{Manini Ojha}
\ead{mojha@jgu.edu.in}
\author[add4]{Mohammad Arshad Rahman\corref{cor1}}
\ead{marshad@iitk.ac.in, arshadrahman25@gmail.com}

\cortext[cor1]{Please address correspondence to Mohammad Arshad
Rahman, Department of Economic Sciences, Indian Institute of Technology, 
Kanpur. Room 672, Faculty Building, IIT Kanpur, Kanpur 208016, India. Phone: 
+91 512-259-7010. Fax: +91 512-259-7500. This book chapter was completed and 
accepted while the corresponding author was working as an Associate Professor 
in the College of Business, Zayed University, UAE.}
\address[add1]{Walmart, India.}
\address[add2]{Department of Economics, Hofstra University, Hempstead, USA}
\address[add3]{Jindal School of Government and Public Policy,
O.P. Jindal Global University, Sonipat, India.}
\address[add4]{Department of Economic Sciences, Indian Institute of
Technology Kanpur, India.}

%
%
%

\begin{abstract}

This chapter presents an overview of a specific form of limited dependent
variable models, namely discrete choice models, where the dependent
(response or outcome) variable takes values which are discrete, inherently
ordered, and characterized by an underlying continuous latent variable.
Within this setting, the dependent variable may take only two discrete
values (such as 0 and 1) giving rise to binary models (e.g., probit and
logit models) or more than two values (say $j=1,2, \ldots, J$, where $J$ is
some integer, typically small) giving rise to ordinal models (e.g., ordinal
probit and ordinal logit models). In these models, the primary goal is to
model the probability of responses/outcomes conditional on the covariates.
We connect the outcomes of a discrete choice model to the random utility
framework in economics, discuss estimation techniques, present the
calculation of covariate effects and measures to assess model fitting. Some
recent advances in discrete data modeling are also discussed. Following the
theoretical review, we utilize the binary and ordinal models to analyze
public opinion on marijuana legalization and the extent of legalization --
a socially relevant but controversial topic in the United States. We obtain
several interesting results including that past use of marijuana, belief
about legalization and political partisanship are important factors that
shape the public opinion.
\end{abstract}

\begin{keyword}
\end{keyword}
\end{frontmatter}

\newpage
\section{Introduction}\label{sec:Intro}

This chapter will discuss settings in which the dependent variable we seek to
model takes on a range of values that are restricted, broadly defined as
\textbf{limited dependent variable models}. Within the class of limited
dependent variables, a special case arises when the outcome is no longer a
continuous measure but a discrete variable. Such data often arise as
individuals making a choice from a set of potential discrete outcomes, thus
earning the name \textbf{discrete choice} models. The most common case of
such models occurs when $y$ is a binary response and takes on the values zero
and one, indicating whether or not the event has occurred, giving rise to
\textbf{binary models}\footnote{Binary models are special cases of both
ordinal and multinomial models with more than two categories.}. Consider for
example, participation in the labour force, whether or not an individual will
buy a vehicle, whether or not a country is part of free trade agreement. In
other cases, $y$ may take on multiple (more than two) discrete values, with
no natural ordering. Consider for example, choice of brand of toothpaste or
mode of transportation. These are referred to as \textbf{multinomial models}.
We refer the readers to \citet{Train-2009} for a detailed discussion on
multinomial models, their estimation and inference. Further, there could be
situations where $y$ takes on multiple (more than two) discrete values that
are inherently ordered or ranked. For example, scores attached to opinion on
surveys (oppose, neutral, support), classification of educational attainment,
or ratings on bonds. These give rise to \textbf{ordinal models} or
\textbf{ordered choice models}. Here, we discuss four discrete choice models
-- ordinal probit, ordinal logit, binary probit and binary logit models.

Discrete choice models have their foundations in the theory of choice in
economics, which itself is inherently related with the random utility model
\citep{Luce-1959, Luce-Suppes-1965, Marschak-1960}. The random utility
framework involves a utility maximizing rational individual whose objective
is to choose an alternative from a set of \textit{mutually exclusive} and
completely \textit{exhaustive} alternatives. The utilities attached with each
alternative are completely known to the decision maker and the agent chooses
the same alternative in replications of the experiment. However, to a
researcher the utilities are unknown, since s/he only observes a vector of
characteristics (such as age, gender, income etc.) of the decision maker,
referred to as \textit{representative utility}. This forms the systematic
component. The unobserved factors form the stochastic part. The stochastic
component is assigned a distribution, typically continuous, to make
probabilistic statements about the observed choices conditional on the
representative utility. The distributional specification implies that there
exists a continuous latent random variable (or a continuous latent utility)
that underlies the discrete outcomes.

When the set of alternatives or outcomes are inherently ordered or ranked,
individual choice of a particular alternative can be associated as the latent
variable crossing a particular threshold or cut-point. This latent variable
threshold-crossing formulation of the ordered choices elegantly connects
individual choice behavior and ordinal data models serve as a useful tool in
the estimation process. While the theoretical support relates to choice and
random utility theory, the econometric techniques are completely general and
applicable when the ordering conditions of the data are met.

To understand the application of discrete choice models, we consider the case
of legalization of marijuana in the United States. The debate around
legalization of marijuana has been an important yet controversial policy
issue. Marijuana has been proved to be effective in treatment of several
diseases and a wealth of new  scientific understanding regarding its
medicinal benefits are documented in \citet{Berman-etal-2004,
Wilsey-etal-2013, Abrams-etal-2003, Abrams-etal-2007, Ellis-etal-2009,
Johnson-etal-2010, McAllister-etal-2011, Guzman-2003,Duran-etal-2010}.
\footnote{The reader is directed to the website www.procon.org for a list of
60 peer-reviewed articles
(\url{http://medicalmarijuana.procon.org/view.resource.php?resourceID=000884})
on the effect of marijuana in treatment of the above mentioned diseases.}
However, despite the medicinal benefits, smoking or consumption of marijuana
is not completely benign and may cause harmful effects, especially associated
with respiratory illnesses and cognitive development \citep{Kalant-2004,
Polen-etal-1993, Meier-etal-2012}.\footnote{A list of peer reviewed articles
on the public health consequences of marijuana can be obtained from the
Office of National Drug Control Policy. Refer to
https://www.whitehouse.gov/ondcp/marijuana.} As a result, several surveys
have been conducted to assess public opinion on the matter. For the purpose
of this chapter, we specifically utilize poll data collected by the Pew
Research Center for the periods 2013 and 2014 to demonstrate the application
of binary models and ordinal models. While there is an increasing trend in
favor of legalizing marijuana based on public opinion, it is noteworthy to
study these specific time periods given that the year 2013 marked the first
time in more than four decades that majority of Americans favored legalizing
the use of marijuana in the United States \citep{Dimock-etal-2013}.

\section{Ordinal Models}\label{sec:Model1}

In \emph{ordinal} regression models, the outcomes of a dependent variable are
categorical and follow a natural ordering. Each outcome or category is
assigned a score (value or number) with the characteristic that the scores
have an ordinal meaning but hold no cardinal interpretation. Therefore, the
difference between categories is not directly comparable. For example, the
study presented in Section~\ref{sec:Study2} codifies the public response to
marijuana legalization as follows: 1 for `oppose legalization', 2 for `legal
only for medicinal use', and 3 for `legal for personal use'. Here, a score of
2 implies more support for legalization as compared to 1, but we cannot
interpret a score of 2 as twice the support compared to a score of 1.
Similarly, the difference in support between 2 and 1 is not the same as that
between 3 and 2.

\begin{figure}[!t]
  \centerline{
    \mbox{\includegraphics[width=7.50in, height=2.25in]{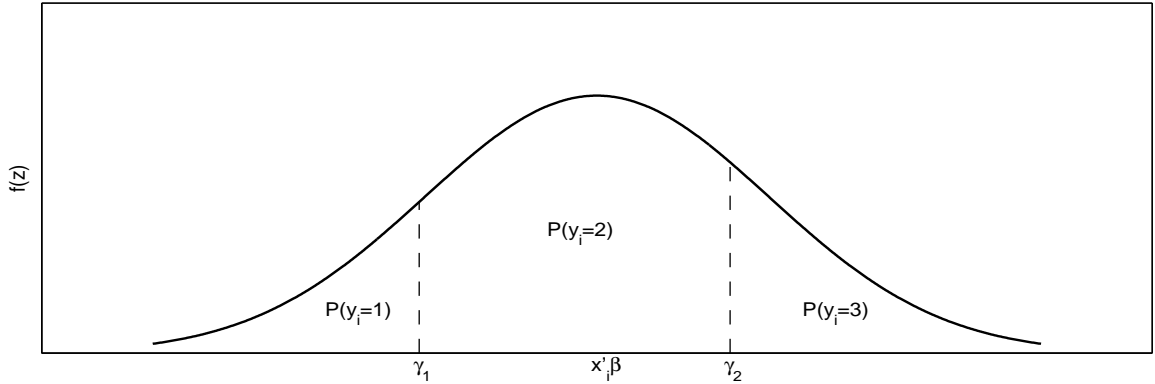}}
    }
\caption{The two cut-points ($\gamma_{1},\gamma_{2}$) divide the
area under the curve into three parts, with each part representing the
probability of a response falling in the three response categories. The three
probabilities P$(y_{i}=1)$, P$(y_{i}=2)$ and P$(y_{i}=3)$ correspond to
`oppose legalization', `legal only for medical use' and `legal for personal
use', respectively. Note that for each individual $i$ the mean $x'_{i}\beta$
will be different and so will be the category probabilities.}
\label{fig:ordinalpdf3}
\end{figure}

While the ordinal regression model has a dependent variable that takes
discrete values, the model can be conveniently expressed in terms of a
continuous latent variable $z_{i}$\footnote{The continuous latent construct
may represent underlying latent utility, some kind of propensity, or strength
of preference \citep{Greene-Hensher-2010}.} as follows:
\begin{equation}
z_{i} = x'_{i} \beta  + \epsilon_{i}, \hspace{0.75in} \forall \; i=1, \cdots, n,
\label{eq:model}
\end{equation}
where $x_{i}$ is a $k \times 1$ vector of covariates, $\beta$ is a $k \times
1$ vector of unknown parameters and $n$ denotes the number of observations.
Like most applications, the stochastic term $\epsilon_{i}$ is assumed to be
\emph{independently and identically distributed} (\emph{iid}) as a standard
normal distribution, i.e., $\epsilon_{i} \mathop\sim\limits^{iid} N(0,1)$ for
$i=1,\ldots,n$; which gives rise to an \emph{ordinal probit model} (also
known as \emph{ordered probit model}). The latent variable $z_{i}$ is related
to the observed discrete response $y_{i}$ as follows:
\begin{equation}
\gamma_{j-1} < z_{i} \leq \gamma_{j} \; \Rightarrow \;
\emph{$y_{i}$ = j}, \hspace{0.75in}
\forall \; i=1,\cdots, n; \; j=1,\cdots, J,
\label{eq:cutpoints}
\end{equation}
where $-\infty = \gamma_{0} < \gamma_{1} \cdots < \gamma_{J-1} < \gamma_{J} =
\infty$ are the cut-points (or thresholds)  and $y_{i}$ is assumed to have
$J$ categories or outcomes. A visual representation of the outcome
probabilities (for the case of marijuana legalization) and the cut-points are
presented in Figure~\ref{fig:ordinalpdf3}. One may observe from
Figure~\ref{fig:ordinalpdf3} that different combinations of $(\beta, \gamma)$
can produce the same outcome probabilities giving rise to parameter
identification problem. We therefore need to anchor the location and scale of
the distribution to identify the model parameters. The former is achieved by
setting $\gamma_{1}=0$ and the latter by assuming
$\textrm{var}(\epsilon_{i})=1$. Other identification schemes are possible and
the reader is referred to \citet{Jeliazkov-etal-2008} and
\citet{Jeliazkov-Rahman-2012} for details.

Given a data vector $y$ = $(y_{1}, \cdots, y_{n})'$, the likelihood for the
ordinal probit model expressed as a function of unknown parameters $(\beta,
\gamma)$ is the following,
\begin{equation}
\begin{split}
\ell(\beta, \gamma; y)
      &  =  \prod_{i=1}^{n} \prod_{j=1}^{J} \Pr(y_{i} = j | \beta,
            \gamma)^{ I(y_{i} = j)},  \\
      &  =  \prod_{i=1}^{n}  \prod_{j=1}^{J}
            \bigg[ \Phi(\gamma_{j} - x'_{i}\beta) -
            \Phi(\gamma_{j-1} - x'_{i}\beta)   \bigg]^{ I(y_{i} = j)},
            \label{eq:likelihoodOP}
\end{split}
\end{equation}
where $\Phi(\cdot)$ denotes the \emph{cumulative distribution function (cdf)}
of a standard normal distribution and $I(y_{i}=j)$ is an indicator function,
which equals 1 if the condition within parenthesis is true and 0 otherwise.
The parameter estimates for ($\beta, \gamma$) are obtained by maximizing the
logarithm of the likelihood (equation~\ref{eq:likelihoodOP}) using numerical
techniques such as the Newton-Raphson method or BHHH procedure
\citep{Train-2009}. The principle behind maximizing the likelihood -- known
as maximum likelihood (ML) estimation -- is to obtain those parameter values
that are most probable to have produced the data under the assumed
statistical model. Note that it is convenient to work with the logarithm of
the likelihood (log-likelihood) since logarithm being a monotonic function,
the maximum of the log-likelihood and the likelihood occur at the same
parameter values. Once the parameter estimates are available, they may be
used to calculate the covariate effects, make predictions or assess model
fitting. Interested readers may look into \citet{Greene-Hensher-2010} or
\citet[][Chap.~4]{Johnson-Albert-2000} for a detailed review of ordinal data
modeling.

Thus far, we have described the ordinal probit model but the framework can be
transformed into an \emph{ordinal logit model} (or \emph{ordered logit
model}) by simply assuming that the error follows a logistic distribution
\citep{McKelvey-Zavoina-1975,McCullagh-1980}. Therefore, for the model in
equation~\eqref{eq:model}, we now assume that $\epsilon_{i} \sim L(0,1)$ for
$i=1,\ldots, n$, where $L$ denotes a logistic distribution with mean $0$ and
variance $\pi^{2}/3$. Like the normal distribution, the logistic distribution
is symmetric but has heavier tails relative to a normal distribution. The
likelihood for the ordinal logit model has the same structure as
equation~\eqref{eq:likelihoodOP} with $\Phi(w)$ replaced by $\Lambda(w) =
\exp(w)/[1 + \exp(w)]$, where $w$ is the argument inside the parenthesis.
Analogous to the ordinal probit model, the parameters are estimated using the
ML technique.

An interesting property of the ordinal logit model is that the ratio of odds
of not exceeding a certain category (say $j$) for any two individuals is
constant across response categories. This earns it the name
\emph{proportional odds model}. To see this property in effect, let
$\theta_{ij} = \Pr(y_{i} \le j)$ denote the cumulative probability that
individual $i$ chooses category $j$ or below. For the ordinal logit model,
this implies: $\theta_{ij} = \exp{(\gamma_{j} - x'_{i}\beta)}/ \big[1 +
\exp{(\gamma_{j} - x'_{i}\beta )} \big]$, and $ \theta_{ij}/(1-\theta_{ij}) =
\exp \big[\gamma_{j} - x'_{i}\beta \big]$, where the latter represents the
odds for the event $y_{i} \le j$. Accordingly, for any two individuals (say
$1$ and $2$), the ratio of odds is,
\begin{equation}
\frac{\theta_{1j}/(1-\theta_{1j})}{\theta_{2j}/(1-\theta_{2j})} =
\exp\big[ - (x_{1} - x_{2})' \beta  \big].
\label{eq:OddsRatio}
\end{equation}
The odds ratio presented in equation~\eqref{eq:OddsRatio} does not depend on
the response category $j$ and is proportional to $(x_{1} - x_{2})$ with
$\beta$ being the constant of proportionality. Thus, the name
\emph{proportional odds model}.

\section{Binary Models}\label{sec:Model2}

\begin{figure*}[!t]
  \centerline{
    \mbox{\includegraphics[width=7.50in, height=2.25in]{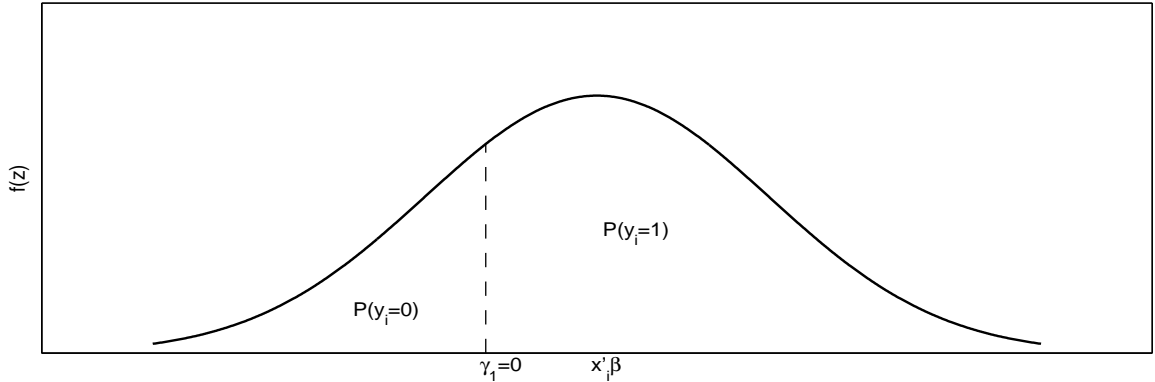}}
    }
\caption{The cut-point $\gamma_{1}$ divides the area under the curve into two
parts, the probability of failure and probability of success. In our study,
P$(y_{i}=0)$ and P$(y_{i}=1)$ correspond to probability of
opposing and supporting marijuana legalization, respectively. Note that for
each individual $i$ the mean $x'_{i}\beta$ and hence the probabilities,
P$(y_{i}=0)$ and P$(y_{i}=1)$, will be different.}
\label{fig:ordinalpdf2}
\end{figure*}

\emph{Binary} models are a simplification of ordinal models and are designed
to deal with situations where the outcome (response or dependent) variable is
dichotomous i.e., takes only two values, typically coded as 1 for `success'
and 0 for `failure'. For example, the application presented in
Section~\ref{subsec:Study1} models the response as a `success' if an opinion
is in favor of legalization and a `failure' otherwise. The general set up of
a binary probit model (or simply probit model) is similar to an ordinal
probit model and can be written in terms of a continuous latent variable
$z_{i}$\footnote{The continuous latent variable can be interpreted as the
difference between utilities from choice $1$ and $0$ i.e., $z_{i} = U_{i1} -
U_{i0}$, where $U$ denotes utility \citep{Jeliazkov-Rahman-2012}.} as
follows,
\begin{equation}
\begin{split}
z_{i} & =  x'_{i} \beta  + \epsilon_{i}, \hspace{0.75in} \forall \;
i=1, \cdots, n, \\
y_{i} & = \left\{ \begin{array}{ll}
1 & \textrm{if} \;  z_{i} > 0,\\
0 & \textrm{otherwise}.
\end{array} \right.
\end{split}
\label{eq:ModelBinary}
\end{equation}
where $\epsilon_{i} \sim N(0,1)$ for $i=1,\ldots,n$. With only two responses,
there is a single cut-point which is typically fixed at 0 for the sake of
simplicity. A pictorial representation of the binary outcome probabilities
for marijuana legalization is
shown in Figure~\ref{fig:ordinalpdf2}. The figure also shows that the
cut-point $\gamma_{1}$ is fixed at 0 to anchor the location of the
distribution. Besides, the scale is fixed by assuming the variance of the
normal distribution is 1. Both the restrictions, as mentioned earlier, are
necessary to identify the model parameters.

The likelihood for the binary probit model can be expressed as,
\begin{equation}
\begin{split}
\ell(\beta; y)
      &  =  \prod_{i=1}^{n}
            \big\{ \Pr(y_{i} = 0|x'_{i}\beta)^{(1-y_{i})}
            \Pr(y_{i} = 1|x'_{i}\beta)^{y_{i}} \big\}, \\
      &  =  \prod_{i=1}^{n}
            \big\{ \Phi(- x'_{i}\beta)^{(1- y_{i})}
            \Phi(x'_{i}\beta)^{y_{i}} \big\}.
            \label{eq:likelihoodBP}
\end{split}
\end{equation}
Given the likelihood, the model parameters are estimated using the ML
technique i.e., by maximizing the log-likelihood
(equation~\ref{eq:likelihoodBP}) with respect to the parameter vector
$\beta$. Once the parameter estimates are available, we may calculate objects
of interest, such as the covariate effects and predicted probabilities.
Measures for goodness of fit can also be calculated to assess model fitting.
Readers interested in further details about binary data modeling may look
into \citet[][Chap.~3]{Johnson-Albert-2000}.

Similar to ordinal models, the framework for the binary probit model given by
equation~\eqref{eq:ModelBinary} can be utilized to describe a binary logit
model (or simply logit model) with the modification that the errors follow a
logistic distribution \citep{Hosmer-etal-2013}. Both the location and scale
restrictions still apply to the logit model, but note that the variance is
now fixed at $\pi^{2}/3$ as compared to 1 in a probit model. To obtain the
logit likelihood, the normal \emph{cdf} $\Phi(x'_{i}\beta)$ in
equation~\eqref{eq:likelihoodBP} is replaced by the logistic \emph{cdf}:
$\Lambda (x'_{i}\beta) = \exp{(x'_{i}\beta)}/ \big[1 + \exp{(x'_{i}\beta )}
\big] $. We can maximize the resulting log-likelihood to obtain the parameter
estimates for the logit model.

The logit model is appealing to researchers in many fields (including
epidemiologists) because of the ease in interpreting its slope coefficient.
To see this, let $\theta_{i} = \Pr(y_{i}=1|x'_{i}\beta)$ denote the
probability of success and $x_{.,l}$ be a continuous covariate (or
independent variable). Then the logarithm of the odds (log-odds) of success
can be expressed as,
\begin{equation*}
\log \bigg( \frac{\theta_{i}}{1 - \theta_{i}}  \bigg) = x'_{i} \beta
= x_{i,l} \, \beta_{l} + x'_{i,-l} \, \beta_{-l}.
\end{equation*}
where $x_{i} = (x_{i,l}, \, x_{i,-l} )$, $\beta = (\beta_{l}, \,
\beta_{-l})$, and $-l$ in the subscript denotes all covariates/parameters
except the $l$-th covariate/parameter. If we differentiate the log-odds with
respect to the $l$-th covariate, we obtain $\beta_{l}$. Therefore, the slope
coefficient $\beta_{l}$ represents the log-odds for a 1 unit change in the
$l$-th covariate.

Similarly, the coefficient of an indicator variable (dummy or dichotomous
variable) has an interesting interpretation. Let $x_{.,m}$ be an indicator
variable, $\theta_{i}^{1}$ be the probability of success when $x_{i,m}=1$,
$\theta_{i}^{0}$ be the probability of success when $x_{i,m}=0$. Our goal is
to find the expression for the odds-ratio, which measures the odds of success
among those with $x_{i,m}=1$ compared to those with $x_{i,m}=0$. Then the
logarithm of the odds-ratio is,
\begin{equation*}
\log \bigg( \frac{\theta_{i}^{1}/(1 - \theta_{i}^{1})}
{\theta_{i}^{0}/(1 - \theta_{i}^{0})}
\bigg) = \beta_{m} + x'_{i,-m} \, \beta_{-m} - x'_{i,-m} \, \beta_{-m}
= \beta_{m}.
\end{equation*}
The odds-ratio is better understood with the help of an example. Suppose $y$
denotes the presence or absence of a heart disease and $x_{.,m}$ denotes
whether the person is a smoker or non-smoker. Then, an odds-ratio $=2$ implies
that heart disease is twice as likely to occur among smokers as compared to
non-smokers for the population under study.

\section{Covariate Effects and Model Fitting}\label{sec:CEMF}

In ordinal models, the coefficients do not give covariate effects because the
link function is non-linear and non-monotonic. Consequently, we need to
calculate the covariate effect for each outcome. Let $x_{.,l}$ be a
continuous covariate, then covariate effect for the $i$-th observation (or
individual) in an ordinal probit model is calculated as,
\begin{equation}
\begin{split}
\frac{\partial \Pr(y_{i}=j)}{\partial x_{i,l}} & = -\beta_{l} \;
\big[ \phi(\gamma_{j} - x'_{i}\beta) - \phi(\gamma_{j-1}-x'_{i}\beta) \big] \\
& \simeq  -\hat{\beta}_{l} \;
\big[ \phi(\hat{\gamma}_{j} - x'_{i}\hat{\beta})
- \phi(\hat{\gamma}_{j-1}-x'_{i}\hat{\beta}) \big],
\end{split}
\label{eq:CEContOrdinal}
\end{equation}
where $\phi(\cdot)$ denotes the probability density function (\emph{pdf}) of
a standard normal distribution and $(\hat{\beta}, \hat{\gamma})$ are the ML
estimates of the parameters $(\beta, \gamma)$. The average covariate effect
is computed by averaging the covariate effect in
equation~\eqref{eq:CEContOrdinal} across all observations. If the covariate
is an indicator variable (say $x_{.,m})$, then the covariate effect for the
$i$-th observation on outcome $j$ ($=1,\ldots,J$) is calculated as,
\begin{equation}
\begin{split}
& \Pr(y_{i}=j|x_{i,-m}, x_{i,m}=1) - \Pr(y_{i}=j|x_{i,-m}, x_{i,m}=0) \\
& =  \big[ \Phi(\gamma_{j} - {x'}^{\dagger}_{i}\beta) - \Phi(\gamma_{j-1}
     - {x'}^{\dagger}_{i}\beta)\big] -
     \big[ \Phi(\gamma_{j-1} - {x'}^{\ddagger}_{i}\beta)
     - \Phi(\gamma_{j-1} - {x'}^{\ddagger}_{i}\beta)\big]\\
& \simeq \big[ \Phi(\hat{\gamma}_{j} - {x'}^{\dagger}_{i}\hat{\beta})
     - \Phi(\hat{\gamma}_{j-1} - {x'}^{\dagger}_{i} \hat{\beta})\big] -
     \big[ \Phi(\hat{\gamma}_{j-1} - {x'}^{\ddagger}_{i} \hat{\beta})
     - \Phi(\hat{\gamma}_{j-1} - {x'}^{\ddagger}_{i}\hat{\beta})\big],
\end{split}
\label{eq:CEIndOrdinal}
\end{equation}
where ${x}^{\dagger}_{i} = (x_{i,-m}, \, x_{i,m}=1)$ and ${x}^{\ddagger}_{i}
= (x_{i,-m}, \, x_{i,m}=0)$. The average covariate effect is calculated by
averaging the covariate effect given in equation~\eqref{eq:CEIndOrdinal}
across all observations. Note that for ordinal models, the sign of the
regression coefficient translates unambiguously into the sign of covariate
effect only for the lowest and highest categories of the response variable.
Covariate effect for the middle categories cannot be known \emph{a priori}.

Moving on to binary probit model, the expressions for the covariate effects
simplify. For a continuous variable, the covariate effect is given by the
expression,
\begin{equation}
\frac{\partial \Pr(y_{i}=1)}{\partial x_{i,l}}  = \beta_{l} \,
\phi(x'_{i}\beta) \simeq \hat{\beta}_{l} \; \phi(x'_{i} \hat{\beta}),
\label{eq:CEContBinary}
\end{equation}
and the same for an indicator variable is given by the expression,
\begin{equation}
\begin{split}
& \Pr(y_{i}=1|x_{i,-m},x_{i,m}=1) - \Pr(y_{i}=1|x_{i,-m}, x_{i,m}=0) \\
& = \Phi({x'}^{\dagger}_{i}\beta) - \Phi({x'}^{\ddagger}_{i}\beta)
\simeq \Phi({x'}^{\dagger}_{i}\hat{\beta}) - \Phi({x'}^{\ddagger}_{i}{\beta}),
\end{split}
\label{eq:CEIndBinary}
\end{equation}
where all the notations have been explained in the previous paragraph. Once
again, the average covariate effect is computed by averaging across all
observations. While the discussion on covariate effects has considered
ordinal and binary probit models because of their implementation in the
applications, covariate effects for the ordinal and binary logit models can
be calculated analogously by replacing the normal \emph{pdf}'s and
\emph{cdf}'s with logistic \emph{pdf}'s and \emph{cdf}'s at appropriate
places.

To assess the goodness of model fit, we calculate three measures: likelihood
ratio (LR) test statistic, McFadden's R-square \citep{McFadden-1974} and
hit-rate \citep{Johnson-Albert-2000}. For the null hypothesis $H_{0}:
\beta_{2} = \ldots = \beta_{k}=0$, the LR test statistic $\lambda_{LR}$ is
defined as follows:
\begin{equation*}
  \lambda_{LR} = - 2 [\ln L_{0} - \ln L_{\mathrm{fit}}] \quad
  \mathop\sim\limits^{H_{0}} \quad \chi^{2}_{k-1}
\end{equation*}
where $\ln L_{\mathrm{fit}}$ is the log-likelihood of the fitted model and
$\ln L_{0}$ is the log-likelihood of the intercept-only model. Under the null
hypothesis, $\lambda_{LR}$ follows a chi-square distribution with degrees of
freedom equal to $k-1$, i.e., the number of restrictions under the null
hypothesis. So, we calculate the statistic $\lambda_{LR}$ and compare it with
$\chi^{2}_{k-1}$ for a given level of significance. If $\lambda_{LR} >
\chi^{2}_{k-1}$, then we reject the null hypothesis. Otherwise, we do not
reject the null hypothesis.

Another popular goodness of fit measure for discrete choice models is
McFadden's R-square ($R^{2}_{M}$), due to \citet{McFadden-1974}. The
McFadden's R-square, also referred to as pseudo R-square or likelihood ratio
index, is defined as follows,
\begin{equation*}
  R^{2}_{M} = 1 - \frac{\ln L_{\mathrm{fit}}}{\ln L_{0}}
\end{equation*}
The $R^{2}_{M}$ is intuitively appealing because it is bounded between 0 and
1, similar to the coefficient of determination $(R^{2})$ in linear regression
models. When all slope coefficients are zero, the $R^{2}_{M}$ equals zero;
but in discrete choice models $R^{2}_{M}$ can never equal 1, although it can
come close to 1. While higher values of $R^{2}_{M}$ implies better
fit, the value as such has no natural interpretation in sharp contrast to
$R^{2}$ which denotes the proportion of variation in the dependent variable
explained by the covariates.

While both LR test statistic and McFadden's R-square are commonly used in
applied studies, the hit-rate is relatively uncommon. The hit-rate (HR) is
defined as the percentage of correct predictions i.e., percentage of
observations for which the model correctly assigns the highest probability to
the observed response category. Mathematically, the HR can be defined as
follows,
\begin{equation*}
HR = \frac{1}{n} \sum_{i=1}^{n} I\bigg( \Big(
\max_{j} \; \{ \hat{p}_{ij} \}_{j=1}^{J} \Big) = y_{i}   \bigg),
\end{equation*}
where $\hat{p}_{ij}$ is the predicted probability that individual $i$ selects
outcome $j$, and $I(\cdot)$ is the indicator function as defined earlier.

\section{Some Advances in Discrete Choice Modeling}\label{sec:SomeAdv}

Till now, we have looked at ordinal and binary models and their Classical (or
Frequentist) approach to estimation via the maximum likelihood technique,
which only involves the likelihood function (say $f(y|\theta)$, where
$\theta$ is the parameter vector). The Classical approach assumes that model
parameters are unknown but have fixed values and hence the parameters cannot
be treated as random variables. In contrast, Bayesian approach to estimation
utilizes the Bayes' theorem,
\begin{equation*}
\pi(\theta|y) = \frac{f(y|\theta) \pi(\theta)}{\int f(y|\theta) \pi(\theta)
\; d\theta} \; ,
\label{eq:BayesTheorem}
\end{equation*}
to update the belief/information about $\theta$ (considered a random
variable) by combining information from the observed sample (via the
likelihood function) and non-sample or prior beliefs (arising from previous
studies, theoretical consideration, researcher's belief, etc.) represented by
the prior distribution $\pi(\theta)$. Inference is based on the posterior
distribution $\pi(\theta|y)$. Bayesian approach provides several advantages
including finite sample inference, working with likelihoods which are
difficult to evaluate, and advantages in computation. Interested readers may
look into \citet{Greenberg-2012} for details on the Bayesian approach and
\citet{PoirierBook-1995} for a comparison of Classical and Bayesian
estimation methods.

The posterior density for ordinal and binary models do not have a tractable
density and so the parameters cannot be sampled directly. While the
Metropolis-Hastings (MH) algorithm \citep{Metropolis-etal-1953,
Hastings-1970} can be employed to sample the parameters, the standard and
more convenient approach is to consider \emph{data augmentation}
\citep{Tanner-Wong-1987}. In this approach, the joint posterior density is
augmented by a latent variable $z$ and the augmented joint posterior
$\pi(\theta,z|y)$ is written as,
\begin{equation*}
\pi(\theta,z|y) \propto \pi(\theta) f(z,y|\theta) =
\pi(\theta) f(z|\theta) f(y|z,\theta).
\label{eq:AugJointPost}
\end{equation*}
For an ordinal probit model, $\theta = (\beta, \gamma)$ and $f(y|z,\beta,
\gamma) \equiv f(y|z,\gamma)$; whereas for a binary probit model $\theta =
\beta$ and $f(y|z,\beta) \equiv f(y|z)$. The two equivalencies arise because,
given a latent observation $z_{i}$, $y_{i}$ is known with certainty
regardless of $\beta$ for $i=1,\ldots,n$. This de-linking of the likelihood
function from $\beta$, made possible through data augmentation, simplifies
the estimation procedure and allows sampling of $\beta$ through a Gibbs
process \citep{Geman-Geman-1984} -- a well known Markov chain Monte Carlo
(MCMC) technique. The latent variable $z$ is sampled element-wise from a
truncated normal distribution. For ordinal models, a monotone transformation
of the cut-points, $\gamma$, is sampled using an MH algorithm. The MCMC
algorithms for estimating ordinal and binary probit models outlined
here were introduced in \citet{Albert-Chib-1993}. Other notable references
that describe the Bayesian modeling and estimation of ordinal and binary
responses in great detail include \citet{Johnson-Albert-2000},
\citet{Greenberg-2012}, and \citet{Jeliazkov-Rahman-2012}. Bayesian
estimation of logit model is based on the same principle and presented in
\citet{Holmes-Held-2006} and \citet{Jeliazkov-Rahman-2012}.

The ordinal and binary models considered in this chapter, whether estimated
using the Classical or the Bayesian techniques, provide information on the 
average probability of outcomes conditional on the covariates. However, 
interests in the quantiles of the response variable as a robust alternative 
to mean regression have grown enormously since the introduction of quantile
regression in \citet{Koenker-Basset-1978}. Quantile modeling gained further
momentum with the development of Bayesian quantile regression by
\citet{Yu-Moyeed-2001}, where the authors create a working likelihood by
assuming that the errors follow an asymmetric Laplace (AL) distribution
\citep{Yu-Zhang-2005}. Binary quantile regression was proposed by
\citet{Kordas-2006} and its Bayesian formulation was presented by
\citet{Benoit-Poel-2010}. \citet{Rahman-2016} introduced Bayesian quantile
regression with ordinal responses and estimated the model using MCMC
techniques. The corresponding R package \emph{bqror} for estimating the 
quantile ordinal model is described in \citet{Maheshwari-Rahman-2023} along 
with the computation of marginal likelihood for comparing alternative 
quantile models. A flexible form of Bayesian ordinal quantile regression was
proposed in \citet{Rahman-Karnawat-2019}. Some recent research on ordinal and
binary quantile regression in the panel/longitudinal set up include
\citet{Rahman-Vossmeyer-2019}, and
\citet{Bresson-etal-2021}. Two applied studies employing ordinal and binary
quantile framework are \citet{Omata-etal-2017} and \citet{Ojha-Rahman-2021},
respectively. Interested readers may explore the above mentioned papers and
references therein to develop a thorough understanding on ordinal and binary
quantile modeling and their applications.

\section{Application: Public Opinion on Legalization of Marijuana in the
United States}

\begin{figure}[!t]
  \centerline{
    \mbox{\includegraphics[width=6.50in, height=3.75in]{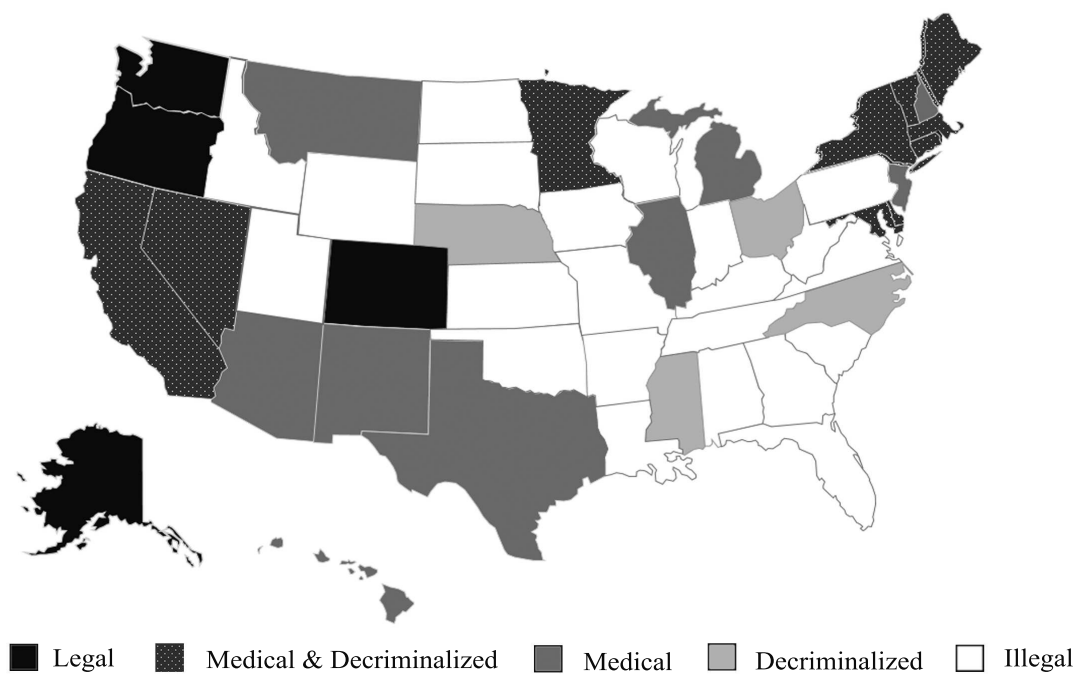}}
    }
\caption{Marijuana state laws in the US as of 6th January, 2016.}
\label{fig:USmap}
\end{figure}

\begin{figure}[!t]
  \centerline{
    \mbox{\includegraphics[width=6.50in, height=3.00in]{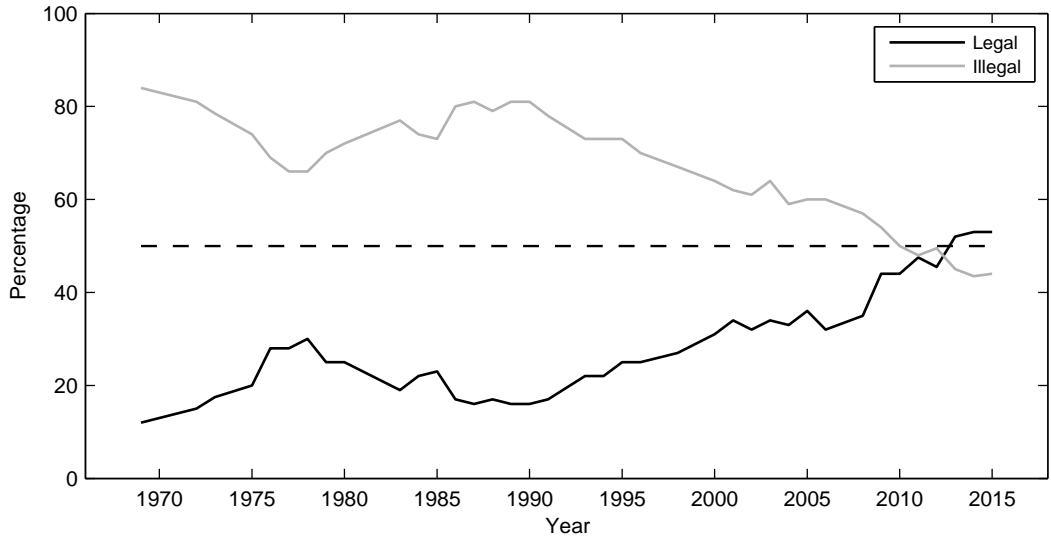}}
    }
\caption{Public opinion on marijuana legalization for the period 1969-2015.
The black dashed line is the 50 percent benchmark. Data source: Pew Research
Center, General Social Survey and Gallup. We have averaged the percentages
for years with multiple surveys. The combined percentage of the two opinions
is below 100 since on average 4 percent of respondents answered
``don't know'' or ``refused to answer''.}
\label{fig:Trend}
\end{figure}

In the United States (US), marijuana is illegal under the federal law as per
the Controlled Substances Act of 1970. The Act classifies marijuana as a
schedule I drug i.e., a drug with no accepted medical value, high potential
for abuse and not safe to use even under medical supervision
\citep{DEA-2011}. However, state laws pertaining to marijuana have evolved
overtime. Up until 2016, 29 states had either legalized, allowed access for
medical reasons or decriminalized its use (See Figure~\ref{fig:USmap}). More
legalization efforts are appearing in the remaining states of US. Such
revisions in state laws represent a change in public attitude that is aptly
reflected in survey data collected by independent polling agencies such as
Pew Research Center, General Social Survey and Gallup. Figure~\ref{fig:Trend}
shows an increasing trend in favor of legalizing marijuana based on public
opinion. Besides, political standing on marijuana has also popularized the
debate on its legalization and may have affected public opinion regarding it.
The gradual growth in support of marijuana perhaps echoes a better public
insight on the medicinal value of marijuana \citep{EarleywineBook-2005} and
the social cost of prohibition that includes illegal trade, racially skewed
arrests of African Americans and huge enforcement cost
\citep{Shepard-Blackley-2007}.


While policies on marijuana use are in the early stages of formulation as
states evaluate its costs and benefits \citep{Winterbourne-2012}, more states
decriminalizing its use may cause a major policy shift at the federal level
\citep{FernerNews-2015}. In this regard, some scholars argue that public
policies ought to be guided by public opinion such that mass opinion and
democracy is upheld \citep{Monroe-1998,Paletz-etal-2015}. Besides,
\citet{Shapiro-2011} cites a large number of studies to argue that public
opinion influences government policy making in the US. Therefore, it is
imperative to study and identify the factors that significantly impact US
public opinion towards marijuana legalization.

In the next section, we employ a binary probit model to analyze public
opinion on marijuana legalization and thereafter implement the ordinal probit
model to analyze public opinion on the extent of marijuana legalization. The
choice of probit models over their logit counterparts is driven by practical
considerations -- the probit models are tractable in univariate cases and can
be generalized to multivariate and hierarchical settings
\citep{Jeliazkov-etal-2008}. In contrast, logistic model based on logistic
distribution cannot model correlations in multivariate settings.

\subsection{Binary Probit Model}\label{subsec:Study1}

\subsubsection{Data}\label{subsubsec:bpData}

We utilize the March 2013 Political Survey data from the Pew Research Center
to analyze public opinion on marijuana and identify the factors that
significantly impact the probability of supporting its legalization. The
survey was conducted during the period March 13-17, 2013, by Abt SRBI
(Schulman, Ronca \& Buculvas, Inc) for the Pew Research Center for the People
and the Press. The survey selected and interviewed a representative sample of
1,501 adults living in the US. Of the 1501 adults, 750 individuals were
interviewed over land line and the remaining 751 individuals over cell phone.
The available sample had several respondents with missing values (``don't
know'' or ``refused to answer'') on the variables of interest, along with 49
respondents who were unsure about marijuana legalization. After removing data
on these respondents, we have a sample of 1182 observations available for the
study.

\begin{table}[!t]
\centering \small \setlength{\tabcolsep}{8pt} \setlength{\extrarowheight}{2pt}
\setlength\arrayrulewidth{1pt}
\caption{Descriptive summary of the variables (March 2013 Political Survey).}\label{Table:Summary1}
\begin{tabular}{llr r}
\toprule
\textsc{variable}       &            & \textsc{mean} & \textsc{std}  \\
\midrule
\textsc{log age}        &              &    3.86   &   0.40    \\
\textsc{log income}     &              &   10.64   &   0.98    \\
\textsc{household size} &              &    2.72   &   1.44    \\
                    \midrule
                    &  \textsc{category}         & \textsc{counts} & \textsc{percentage} \\
                    \midrule
\textsc{past use}   &                  &   554     &   46.87      \\
\textsc{male}       &                  &   570     &   48.22      \\
                    \cmidrule{2-2}
                    & \textsc{bachelors \& above}
                                       &   426     &   36.04      \\
\textsc{education}  & \textsc{below bachelors}
                                       &   360     &   30.46      \\
                    & \textsc{high school \& below}
                                       &   396     &   33.50      \\
                    \cmidrule{2-2}
\textsc{tolerant states}   &           &   374     &   31.64      \\
                    \cmidrule{2-2}
                    & \textsc{white}   &   938     &   79.36      \\
\textsc{race}
                    & \textsc{african american}
                                       &   142     &   12.01      \\
                    & \textsc{other races}
                                       &   102     &    8.63      \\
                    \cmidrule{2-2}
                    & \textsc{republican}
                                       &   353     &   29.86      \\
\textsc{party affiliation}
                    & \textsc{democrat}&   404     &   34.18      \\

                    & \textsc{independent \& others}
                                       &   425     &   35.96      \\
                    \cmidrule{2-2}
                    & \textsc{protestant}
                                       &   494     &   41.79      \\
                    & \textsc{roman catholic}
                                       &   258     &   21.83      \\
\textsc{religion}
                    & \textsc{christian}
                                       &   138     &   11.68      \\
                    & \textsc{conservative}
                                       &    72     &    6.09      \\
                    & \textsc{liberal}
                                       &   220     &   18.61      \\
\midrule \textsc{public opinion}
                    & \textsc{favor legalization}
                                       &   622     &   52.62      \\
                    & \textsc{oppose legalization}
                                       &   560     &   47.38      \\
\bottomrule
\end{tabular}
\end{table}

In this application, the dependent variable is response to the question: ``Do
you think the use of marijuana should be made legal, or not?''. The responses
were recorded as `yes, legal' (i.e. favor legalization), `no, illegal' (i.e.,
oppose legalization) or `don't know or refused'. We remove the last category
as it constitutes missing responses. This makes the response variable binary
and hence a binary probit model is utilized to analyze the response based on
the following set of covariates: age, income, household size, past use of
marijuana, gender, education, state of residence, race, party affiliation and
religion.

Age was recorded in years. Income (measured in US Dollars) was reported as
belonging to one of 9 groups (0-10k, 10k-20k, $\cdots$, 40k-50k, 50k-75k,
75k-100k, 100k-150k, 150k and above, where `k' denotes thousand). We convert
income to a continuous variable by taking the mid-point of the first 8 income
groups and impute 150k for the last group. Household size represents the
number of members in the family. Past use of marijuana and gender are
indicator variables in the model. Educational attainment of the respondents
is classified into three categories
and the category `high school and below' forms the base or reference category
in the regressions. The variable `tolerant states' indicates if a respondent
lives in one of the 20 states, where recreational usage is legal, possession
is decriminalized and/or allowed for medical use only.\footnote{The list of
tolerant states before the date of the survey include Alaska, Arizona,
California, Colorado, Connecticut, Delaware, Hawaii, Maine, Maryland,
Massachusetts, Michigan, Montana, Nevada, New Jersey, New Mexico, Oregon,
Rhode Island, Vermont, Washington, Washington DC. Source:
https://www.whitehouse.gov/ondcp/state-laws-related-to-marijuana.} Race is
classified into three categories and
`White' race is used as the base category in the regressions. The category
`Other Races' comprises of Asian, Hispanic, native American, Pacific
Islanders and remaining races. Party affiliation is also classified into
three categories and Republican Party is used as the reference category in
the models. Religion is classified into five categories and `Protestant' is
used as the base category. Here, the category `Conservative' comprises of
respondents belonging to one of the following religions: Buddhist, Hindu,
Islam, Jew, Mormon and orthodox church. The category `Liberal' comprises of
respondents who claim to be Agnostic, Atheist, Universalist or nothing in
particular. The descriptive statistics for all the variables are presented in
Table~\ref{Table:Summary1}.

Let us now look at the socio-demographic characteristics of a typical
respondent in the sample. An average respondent is about 51 years old and
s/he belongs to a household of size 3 with an annual income of 60,527 US
Dollars. The sample is almost evenly split between males and females and
46.87 percent of respondents have a history of marijuana use. In the sample,
the largest proportion of respondents (36.04 percent) have `bachelors and
above' degree, followed by `below bachelors' degree (30.46 percent). A
significant fraction of the respondents (31.64 percent) reside in states that
have some favorable laws towards marijuana. The sample is predominantly White
(79.36 percent) with a good representation (12.01 percent) of the African
American population. Amongst the respondents, almost 30 percent consider
themselves as Republican, about 34 percent declare themselves as Democrats
and the remaining are Independent or belong to other parties. With respect to
religious codification, the largest proportion (41.79 percent) is
Protestants, followed by Roman Catholics (21.83 percent). A good proportion,
11.68 percent, declare themselves to be simply Christian. The Liberal
category forms about 18.61 percent and the Conservatives have the lowest
fraction at 6.09 percent.

\subsubsection{Estimation }\label{subsubsec:bpEstimation}

We estimate four different binary probit models and present the estimated
coefficients and standard errors for each model in
Table~\ref{Table:BinaryModelResults}. To begin with, Model~1 considers a
basic set of covariates that includes log age, income, past use of marijuana,
gender, education categories, household size and state of residence of the
respondents. Subsequent models generalize Model~1 by adding more variables to
the basic set of regressors. Specifically, Model~2 adds the race variable,
Model~3 adds party affiliation to the list of variables in Model~2, and
Model~4 incorporates religious denomination to the regressors in Model~3. All
the four models have high LR statistic as shown in
Table~\ref{Table:BinaryModelResults} and hence we reject the null hypothesis
that the coefficients are jointly zero in each model. The two other goodness
of fit statistics, McFadden's $R^{2}$ and hit-rate, also show that all
models provide a good fit, with each subsequent model providing a better fit
than the previous model.

\subsubsection{Results}\label{subsubsec:bpResults}
We focus on the results from Model~4, since it provides the best model fit.
Table~\ref{Table:BinaryModelResults} shows that log age has a negative
coefficient and is statistically significant at 5 percent level.\footnote{The
default significance level is 5 percent and henceforth reference to
significance level will be omitted.} This implies that young people are more
supportive of marijuana legalization. This is not surprising since risk
taking or deviant behavior is high amongst the younger population and is well
documented in the literature \citep{Brown-etal-1974}. Moreover,
\citet{Saieva-2008} reports a negative association between age and
probability of supporting legalization, while \citet{Alfonso-Dunn-2007} and
\citet{Delforterie-etal-2015} note that marijuana prevalence rate is higher
among the younger population.

\begin{table}[!t]
\centering \small \setlength{\tabcolsep}{4pt} \setlength{\extrarowheight}{2pt}
\setlength\arrayrulewidth{1pt}\caption{Estimation results for the binary probit model.}
\begin{tabular}{ll d{3.4} d{1.2} r
                   d{3.4} d{1.2} r
                   d{3.4} d{1.2} r
                   d{3.4} d{1.2} }
\toprule
&&  \multicolumn{2}{c}{\textsc{model 1}}  & &
    \multicolumn{2}{c}{\textsc{model 2}}  & &
    \multicolumn{2}{c}{\textsc{model 3}}  & &
    \multicolumn{2}{c}{\textsc{model 4}}    \\
\cmidrule{3-4} \cmidrule{6-7}  \cmidrule{9-10}  \cmidrule{12-13}
    &&  \multicolumn{1}{c}{\textsc{coef}}  & \multicolumn{1}{c}{\textsc{se}} &&
        \multicolumn{1}{c}{\textsc{coef}}  & \multicolumn{1}{c}{\textsc{se}} &&
        \multicolumn{1}{c}{\textsc{coef}}  & \multicolumn{1}{c}{\textsc{se}} &&
        \multicolumn{1}{c}{\textsc{coef}}  & \multicolumn{1}{c}{\textsc{se}}   \\
\midrule
\textsc{intercept}          &&  1.74^{**}   &  0.61   &&    1.70^{**}   &   0.64         &&
                                0.11        &  0.66   &&    0.51        &   0.69  \\
\textsc{log age}            && -0.41^{**}   &  0.11   &&   -0.41^{**}   &   0.11         &&
                               -0.39^{**}   &  0.11   &&   -0.29^{**}   &   0.12  \\
\textsc{log income}         && -0.06        &  0.05   &&   -0.06        &   0.05         &&
                               -0.02        &  0.05   &&   -0.03        &   0.05  \\
\textsc{past use}           &&  0.81^{**}   &  0.08   &&    0.82^{**}   &   0.08         &&
                                0.82^{**}   &  0.08   &&    0.81^{**}   &   0.08  \\
\textsc{male}               &&  0.18^{**}   &  0.08   &&    0.17^{**}   &   0.08         &&
                                0.21^{**}   &  0.08   &&    0.17^{**}   &   0.08  \\
\textsc{bachelors \& above} &&  0.23^{**}   &  0.10   &&    0.23^{**}   &   0.10         &&
                                0.23^{**}   &  0.11   &&    0.21^{*}    &   0.11  \\
\textsc{below bachelors}    &&  0.19^{*}    &  0.10   &&    0.19^{*}    &   0.10         &&
                                0.24^{**}   &  0.10   &&    0.26^{**}   &   0.10  \\
\textsc{household size}     && -0.04        &  0.03   &&   -0.05        &   0.03         &&
                               -0.04        &  0.03   &&   -0.03        &   0.03  \\
\textsc{tolerant states}    &&  0.13        &  0.08   &&    0.12        &   0.08         &&
                                0.12        &  0.08   &&    0.08        &   0.09  \\
\textsc{african american}   &&   ..         &  ..     &&   -0.05        &   0.12         &&
                               -0.26^{**}   &  0.13   &&   -0.14        &   0.13  \\
\textsc{other races}        &&   ..         &  ..     &&    0.12        &   0.14         &&
                                0.03        &  0.14   &&    0.02        &   0.15  \\
\textsc{democrat}           &&   ..         &  ..     &&    ..          &    ..          &&
                                0.68^{**}   &  0.10   &&    0.57^{**}   &   0.11  \\
\textsc{other parties}      &&   ..         &  ..     &&    ..          &    ..          &&
                                0.48^{**}   &  0.10   &&    0.40^{**}   &   0.10  \\
\textsc{roman catholic}     &&   ..         &  ..     &&    ..          &    ..          &&
                                 ..         &  ..     &&    0.18^{*}    &   0.10  \\
\textsc{christian}          &&   ..         &  ..     &&    ..          &    ..          &&
                                 ..         &  ..     &&   -0.06        &   0.13  \\
\textsc{conservative}       &&   ..         &  ..     &&    ..          &    ..          &&
                                            &  ..     &&    0.44^{**}   &   0.18  \\
\textsc{liberal}            &&   ..         &  ..     &&    ..          &    ..          &&
                                 ..         &  ..      &&   0.66^{**}   &   0.12  \\
\midrule
\textsc{LR ($\chi^{2}$) statistic}
                            &&  & 164.37    &&  &  165.38   &&  &   211.44   &&  &  248.55     \\
\textsc{mcfadden's $R^{2}$}
                            &&  &   0.10    &&  &    0.10    &&  &    0.13    &&  &   0.15       \\
\textsc{hit-rate}
                            &&  &  66.67    &&  &   66.58    &&  &   67.51    &&  &  69.03      \\
\bottomrule
\\
\multicolumn{5}{l}{\footnotesize{$\ast \ast$ p $<$ 0.05, $\ast$ p $<$ 0.10}}
\end{tabular}
\label{Table:BinaryModelResults}
\end{table}

The coefficient for income is negative, but statistically insignificant as
also documented in \citet{Nielsen-2010} and \citet{Saieva-2008}. As such, we
do not note a significant relationship between income and probability of
supporting marijuana legalization. This is in disagreement with the
hypothesis that the economically weaker section will support legalization
since marijuana is popular within the lower income group. Past use of
marijuana may drive support for legalization, but this was not controlled
either in \citet{Nielsen-2010} or \citet{Saieva-2008}. Controlling for this
variable in our models, we find that the coefficient for past use is largest
amongst all variables and highly significant. This finding provides support
to the hypothesis that individuals who have used marijuana in the past
strongly favor its legalization and the large coefficient value implies that
past use is an important factor in favoring legalization. Males are more
supportive of legalizing marijuana compared to females and the coefficient is
significant, a result also documented in \citet{Nielsen-2010} and
\citet{Delforterie-etal-2015}. Similarly, \citet{Rodriguez-2015} finds that
boys are more likely to use marijuana during their adolescent years. Since
past use is an important determinant for supporting legalization and
marijuana use is more prevalent among males, it is not surprising to find
that males are more supportive of legalization.

The indicator variables for higher education i.e., `bachelors \& above' and
`below bachelors' have positive coefficients and are statistically
significant (either at 10 or 5 percent significance level) relative to the
base category, `high school \& below'. Thus, higher education leads to
increased support for legalization possibly because a more educated
individual can better understand the costs and medicinal benefits of
marijuana. However, some studies have found that early use of marijuana leads
to lower educational attainment and poor performance in school
\citep{Lynskey-Hall-2000, VanOurs-Williams-2009, Horwood-etal-2010}.
Household size has a negative effect, but the coefficient is not statistically
significant. Similarly, the coefficient for `tolerant states' is positive,
but insignificant. This implies that residing in one of 20 states that offers
some relaxation on marijuana use/offence does not statistically increase the
probability of supporting legalization.

The coefficient for African American and `Other Races', are not statistically
different from the base category, White, with the exception of Model~3. The
negative coefficient for African American, although insignificant, is rather
surprising because one would expect African Americans to support legalization
in order to curtail the large number of marijuana related arrests from the
African American community. In line with this, \citet{Chen-KilleyaJones-2006}
also examine the extent of marijuana use across race and find that marijuana
use is higher among suburban White students compared to their African
American counterparts. Moreover, \citet{Nasim-etal-2007} document the
cultural orientation for African American young women and find that
traditional religious beliefs and practices could be the reason behind less
marijuana usage among African American.

Political affiliation often represents ideological differences towards any
public policy and several poll studies conducted by Pew and Gallup have found
that Republicans (Democrats) are more likely to oppose (favor) legalization
of marijuana. We also arrive at a similar conclusion, with the coefficients
for Democrat and `Other Parties' being positive and significant. This
suggests that individuals having either political orientations are more
supportive of legalization compared to Republicans and the result is
consistent with \citet{Nielsen-2010}. Lastly, we look at the effect of
religious affiliations since religious beliefs sometime acts as a protective
factor against alcohol usage and smoking. The results show that the
coefficient for Roman Catholic is positive and statistically significant at
10 percent, while coefficient for Christian is negative but statistically
insignificant. In contrast, the coefficients for Conservative and Liberal are
both positive and statistically significant and hence both groups are more
supportive of legalization compared to Protestants. However, individual
opinions often do not strictly adhere to religious codes and conducts, so
these results may vary significantly across samples.

\begin{table}[!t]\centering
\small \setlength{\tabcolsep}{6pt} \setlength{\extrarowheight}{2pt}
\caption{Average covariate effects from Model~4.}
\label{Table:CovEffectBM}
\begin{tabular}{ll S[table-format=-1.4] }
\toprule
\textsc{covariate}   & &  \text{$\Delta$P(favor legalization)}   \\
\midrule
\textsc{age, 10 years}        & &  -0.019        \\
\textsc{past use}             & &   0.285        \\
\textsc{male}                 & &   0.057        \\
\textsc{bachelors \& above}   & &   0.068        \\
\textsc{below bachelors}      & &   0.085        \\
\textsc{democrat}             & &   0.188        \\
\textsc{other parties}        & &   0.134        \\
\textsc{roman catholic}       & &   0.058        \\
\textsc{conservative}         & &   0.143        \\
\textsc{liberal}              & &   0.220        \\
\bottomrule
\end{tabular}
\end{table}

The above discussion suggests that the signs of the coefficients, except the
race variables, are consistent with what one would typically expect. However,
the coefficients by themselves do not give the covariate effects (see
Section~\ref{sec:Model1} and \ref{sec:Model2}). Table~\ref{Table:CovEffectBM}
presents the average covariate effects for all significant variables, either
at 5 or 10 percent level. Results show that an increase in age by 10 years
decreases the probability of support by 1.9 percent. The highest positive
impact comes from past use, which shows that an individual who has used
marijuana is 28.5 percent more likely to support legalization relative to
someone who has never used it. Males are 5.7 percent more likely to support
legalization relative to females. Higher education increases the probability
of support and an individual with bachelors or higher degree (below
bachelors) is 6.8 (8.5) percent more likely to support legalization relative
to an individual with a high school degree or below. Political affiliation to
the Democratic Party increases the probability of support by 18.8 percent.
Similarly, an individual who identifies themselves with Independent and other
parties is 13.4 percent more likely to support legalization compared to a
Republican. Finally, an individual who is
 Conservative (Liberal) is 14.3 (22)
percent more likely to support legalization relative to a Protestant.

\subsection{Ordinal Probit Model - The Extent of Marijuana Legalization}\label{sec:Study2}

The year 2013 was the first year in four decades that majority of Americans
favored legalization of marijuana. While its support has grown overtime as
shown in Figure~\ref{fig:Trend}, it is important to distinguish between
different levels of support. This distinction is crucial because support for
personal use of marijuana is stronger than support for its medical use and
has different policy implications. Individuals may opine to support marijuana
for medicinal benefits, but not for personal use. The February 2014 Political
Survey recorded individual response as a three level categorical variable,
which permits use of an ordinal probit model to study the effect of
covariates on public opinion about the extent of legalization.

\subsubsection{Data}\label{subsubsec:opData}

The February 2014 Political Survey was conducted during February 14-23, 2014
by the Princeton Survey Research Associates and sponsored by the Pew Research
Center for the People and the Press. In the survey, a representative sample
of 1,821 adults living in the US were interviewed over telephone with 481
(1,340) individuals interviewed over land line (cell phone, including 786
individuals without a land line phone). The sampled data contain several
missing observations and many respondents were unsure about their opinion on
legalization. As before, we remove data on these respondents and are left with
 a sample of 1,492 observations for the study.

\begin{table}[!t]
\centering \small \setlength{\tabcolsep}{8pt} \setlength{\extrarowheight}{2pt}
\setlength\arrayrulewidth{1pt}
\caption{Descriptive summary of the variables (February 2014 Political Survey).}\label{Table:Summary2}
\begin{tabular}{llr r}
\toprule
\textsc{variable}       &            & \textsc{mean} & \textsc{std}  \\
\midrule
\textsc{log age}        &              &    3.72   &   0.44    \\
\textsc{log income}     &              &   10.63   &   0.98    \\
\textsc{household size} &              &    2.74   &   1.42    \\
                    \midrule
                    &  \textsc{category}         & \textsc{counts} & \textsc{percentage} \\
                    \midrule
\textsc{past use}   &                  &   719     &   48.19      \\
\textsc{male}       &                  &   792     &   53.02      \\
                    \cmidrule{2-2}
                    & \textsc{bachelors \& above}
                                       &   551     &   36.93      \\
\textsc{education}  & \textsc{below bachelors}
                                       &   434     &   29.09      \\
                    & \textsc{high school \& below}
                                       &   507     &   33.98      \\
                    \cmidrule{2-2}
\textsc{tolerant states}   &           &   556     &   37.27      \\
\textsc{eventually legal}  &           & 1,154     &   77.35      \\
                    \cmidrule{2-2}
                    & \textsc{white}   &  1149     &   77.01      \\
\textsc{race}
                    & \textsc{african american}
                                       &   202     &   13.54      \\
                    & \textsc{other races}
                                       &   141     &    9.45      \\
                    \cmidrule{2-2}
                    & \textsc{republican}
                                       &   333     &   22.32      \\
\textsc{party affiliation}
                    & \textsc{democrat}&   511     &   34.25      \\

                    & \textsc{independent \& others}
                                       &   648     &   43.43      \\
                    \cmidrule{2-2}
                    & \textsc{protestant}
                                       &   550     &   36.86      \\
                    & \textsc{roman catholic}
                                       &   290     &   19.44      \\
\textsc{religion}
                    & \textsc{christian}
                                       &   182     &   12.20      \\
                    & \textsc{conservative}
                                       &   122     &    8.18      \\
                    & \textsc{liberal}
                                       &   348     &   23.32      \\
                    \cmidrule{2-2}
                    & \textsc{oppose legalization}
                                       &   218     &   14.61      \\
\textsc{public opinion}
                    & \textsc{legal only for medicinal use}
                                       &   640     &   42.90      \\
                    & \textsc{legal for personal use}
                                       &   634     &   42.49      \\
\bottomrule
\end{tabular}
\end{table}

The dependent variable in the model is the respondents' answer to the
question, ``Which comes closer to your view about the use of marijuana by
adults?''. The options provided were, `It should not be legal,' `It should be
legal only for medicinal use,' or `It should be legal for personal use'. The
fourth category labeled, `Don't know/Refused' is removed from the study.
Similar to the March 2013 Survey, the February 2014 Survey also collected
information on the age, income, household size, past use, gender, education,
race, party affiliation and religion. We use these variables as independent
variables in the models. All the definitions and categories for the variables
remain the same as in Section~\ref{subsubsec:bpData}. We also include the
indicator variable `tolerant states', with the definition modified to include
Illinois and New Hampshire to the previous list of 20 states.\footnote{Note
that marijuana related laws were passed in Illinois and New Hampshire after
the March 2013 Political Survey, but before the February 2014 Political
Survey.} Finally, we include an additional variable, labeled `eventually
legal', for which data was collected only in the February 2014 Survey. This
variable indicates whether respondents expect marijuana to be legal
irrespective of their individual opinion. We present the descriptive
statistics for all the variables in Table~\ref{Table:Summary2}.

Upon exploration of the socio-demographic characteristics of the current
sample, we note that an average respondent is about 45.5 years old and s/he
belongs to a household of size 3 with an annual income of 60,647 US Dollars.
Thus, the typical respondent is about 6 years younger compared to the March
2013 data and approximately has the same household size and income. The
percentage of males is higher in the current sample by 5 percent, but still
close to a fair split between males and females. Similarly, the sample is
almost evenly split between respondents who have used marijuana and those who
have not. The largest proportion of respondents (36.93 percent) have
`bachelors and above' degree, followed by `high school \& below' degree
(33.98 percent). A significant fraction of the respondents (37.27 percent)
reside in states that have some favorable law on marijuana. Looking at the
additional variable `eventually legal', note that 77.35 percent of the
surveyed people expect marijuana to be legal irrespective of their opinion.
Similar to the earlier data, the sample is predominantly White (77.01
percent) with a good representation (13.54 percent) of the African American
population. Party affiliation shows that 34.25 percent of the sample is
comprised of Democrats, 22.32 percent Republicans and the remaining fraction
are `independent \& others'. With respect to religious classifications, the
largest proportion of respondents are Protestant (36.86 percent), followed by
Liberal (23.32 percent) and Roman Catholics (19.44 percent).

\subsubsection{Estimation}\label{subsubsec:opEstimation}

The ordinal probit model results are presented in
Table~\ref{Table:OrdinalModelResults}, which show the coefficient estimates
and standard errors of four different models. The estimation of models follows
a similar sequence as in Table~\ref{Table:BinaryModelResults}. Model~5 is the
base model and contains log age, log income, past use of marijuana, male,
education categories, household size, tolerant states and eventually legal.
Model~6 adds the race categories to Model~5 and Model~7 adds party
affiliation to the list of variables in Model~6. Finally, Model~8 contains
all the variables in Model~7 and religious categories. The goodness of fit
statistics are presented in the last three rows of
Table~\ref{Table:OrdinalModelResults}. The LR statistics are large and each
model fits better than the respective intercept model. The other two
measures, McFadden's $R^{2}$ and hit-rate, show that Model~7 and Model~8
outperform the remaining two models. While Model~8 provides a better fit
compared to Model~7 as per McFadden's $R^{2}$ (0.1187 compared to 0.1254),
Model~7 has a better hit-rate.

\begin{table}[!t]
\centering \small \setlength{\tabcolsep}{4pt} \setlength{\extrarowheight}{2pt}
\setlength\arrayrulewidth{1pt}\caption{Estimation results for the ordinal probit model.}
\begin{tabular}{ll d{3.4} d{1.2} r
                   d{3.4} d{1.2} r
                   d{3.4} d{1.2} r
                   d{3.4} d{1.2} }
\toprule
&&  \multicolumn{2}{c}{\textsc{model 5}}  & &
    \multicolumn{2}{c}{\textsc{model 6}}  & &
    \multicolumn{2}{c}{\textsc{model 7}}  & &
    \multicolumn{2}{c}{\textsc{model 8}}    \\
\cmidrule{3-4} \cmidrule{6-7}  \cmidrule{9-10}  \cmidrule{12-13}
    &&  \multicolumn{1}{c}{\textsc{coef}}  & \multicolumn{1}{c}{\textsc{se}} &&
        \multicolumn{1}{c}{\textsc{coef}}  & \multicolumn{1}{c}{\textsc{se}} &&
        \multicolumn{1}{c}{\textsc{coef}}  & \multicolumn{1}{c}{\textsc{se}} &&
        \multicolumn{1}{c}{\textsc{coef}}  & \multicolumn{1}{c}{\textsc{se}}   \\
\midrule
\textsc{intercept}          &&  1.26^{**}   &  0.44   &&    1.27^{**}   &   0.45         &&
                                0.83^{*}    &  0.46   &&    0.34        &   0.48  \\
\textsc{log age}            && -0.44^{**}   &  0.07   &&   -0.45^{**}   &   0.07         &&
                               -0.45^{**}   &  0.08   &&   -0.35^{**}   &   0.08  \\
\textsc{log income}         &&  0.07^{*}    &  0.03   &&    0.07^{*}    &   0.03         &&
                                0.08^{**}   &  0.03   &&    0.09^{**}   &   0.04  \\
\textsc{past use}           &&  0.74^{**}   &  0.06   &&    0.73^{**}   &   0.06         &&
                                0.71^{**}   &  0.06   &&    0.69^{**}   &   0.06  \\
\textsc{male}               &&  0.06        &  0.06   &&    0.06        &   0.06         &&
                                0.07        &  0.06   &&    0.06        &   0.06  \\
\textsc{bachelors \& above} &&  0.26        &  0.08   &&    0.26^{**}   &   0.08         &&
                                0.25^{**}   &  0.08   &&    0.24^{**}   &   0.08  \\
\textsc{below bachelors}    &&  0.06        &  0.08   &&    0.05^{*}    &   0.08         &&
                                0.05        &  0.08   &&    0.05        &   0.08  \\
\textsc{household size}     && -0.04^{*}    &  0.02   &&   -0.04        &   0.02         &&
                               -0.03        &  0.02   &&   -0.02        &   0.02  \\
\textsc{tolerant states}    &&  0.11^{*}    &  0.06   &&    0.13^{**}   &   0.06         &&
                                0.09        &  0.06   &&    0.07        &   0.07  \\
\textsc{eventually legal}   &&  0.58^{**}   &  0.07   &&    0.58^{**}   &   0.07         &&
                                0.56^{**}   &  0.07   &&    0.57^{**}   &   0.07  \\
\textsc{african american}   &&   ..         &  ..     &&    0.11        &   0.09         &&
                               -0.01        &  0.10   &&    0.03        &   0.10  \\
\textsc{other races}        &&   ..         &  ..     &&   -0.18^{*}    &   0.11         &&
                               -0.26^{**}   &  0.11   &&   -0.27^{**}   &   0.11  \\
\textsc{democrat}           &&   ..         &  ..     &&    ..          &    ..          &&
                                0.48^{**}   &  0.09   &&    0.44^{**}   &   0.09  \\
\textsc{other parties}      &&   ..         &  ..     &&    ..          &    ..          &&
                                0.40^{**}   &  0.08   &&    0.36^{**}   &   0.08  \\
\textsc{roman catholic}     &&   ..         &  ..     &&    ..          &    ..          &&
                                 ..         &  ..     &&    0.10        &   0.09  \\
\textsc{christian}          &&   ..         &  ..     &&    ..          &    ..          &&
                                 ..         &  ..     &&    0.16        &   0.10  \\
\textsc{conservative}       &&   ..         &  ..     &&    ..          &    ..          &&
                                            &  ..     &&    0.09        &   0.12  \\
\textsc{liberal}            &&   ..         &  ..     &&    ..          &    ..          &&
                                 ..         &  ..     &&   0.39^{**}    &   0.09  \\
\textsc{cut-point}          &&  1.43^{**}   &  0.05   &&   1.43^{**}    &   0.05          &&
                                1.45^{**}   &  0.05   &&   1.46^{**}    &   0.05  \\
\midrule
\textsc{LR ($\chi^{2}$) statistic}
                            &&  &  316.27   &&  &   321.47   &&  &   356.99   &&  &   377.02     \\
\textsc{mcfadden's $R^{2}$}
                            &&  &   0.10    &&  &    0.11    &&  &    0.12    &&  &   0.12       \\
\textsc{hit-rate}
                            &&  &  57.77    &&  &   57.44    &&  &   59.11    &&  &   58.91      \\
\bottomrule
\\
\multicolumn{5}{l}{\footnotesize{$\ast \ast$ p $<$ 0.05, $\ast$ p $<$ 0.10}}
\end{tabular}
\label{Table:OrdinalModelResults}
\end{table}

\subsubsection{Results}\label{subsubsec:opResults}

We focus on the results from Model~8 because it is the most general model,
provides the best fit according to McFadden's $R^{2}$ and no variables change
sign. The results indicate that log age has a negative effect on the support
for personal use (third category) and is statistically significant at 5
percent level. Alternatively, log age has a positive effect on opposing
legalization (first category) and is statistically significant. However, the
effect of age on medicinal use only (second category) cannot be determined
\emph{a priori}. Henceforth, we shall only discuss the effect on personal use
and the impact on opposing legalization will be opposite to that of personal
use. As before, the default level of significance used is 5 percent and
further discussion will omit reference to significance level.

We note that the coefficient for log income is positive and statistically
significant. This implies that individuals with higher income are more likely
to support legalization for personal use. Past use of marijuana has a
statistically significant positive effect on the probability of supporting
personal use of marijuana. Moreover, the coefficient for past use is largest
among all the variables, a result which is similar to that obtained in the
binary probit model. Contrary to the finding in
Section~\ref{subsubsec:bpResults}, the coefficient for male is positive, but
is not statistically significant. Thus, the current data do not confirm any
role of gender on public opinion towards marijuana. Higher education is
positively associated with support for personal use of marijuana. However,
only the coefficient for `bachelors degree \& above' is statistically
significant. The coefficients for household size and `tolerant states' are
not significant and conform to the results from the binary probit model. We
find that `eventually legal' has a significant positive effect, indicating
that if an individual expects marijuana to be legal irrespective of his or
her opinion, then s/he is more likely to support legalization for personal
use.

The race variables suggest that opinions of African Americans on personal use
of marijuana are not significantly different compared to the Whites. Such a
similarity of opinions across race was also observed in the binary probit
model. In contrast, `Other Races' has a significant negative coefficient and
is more opposed to legalization as compared to Whites. The coefficients for
political party affiliations are in consonance with the results from
Section~\ref{subsubsec:bpResults}. Affiliation to Democratic Party or `Other
Parties' increases the support for personal use and the coefficients are
significant. Lastly, religious affiliations do not show a strong effect.
Here, only the Liberals are more supportive of personal use of marijuana,
while the opinions of the remaining religious categories are not
significantly different from the base category, Protestant.

\begin{table}[!t]\centering
\small \setlength{\tabcolsep}{6pt} \setlength{\extrarowheight}{2pt}
\caption{Average covariate effects from Model~8.}
\label{Table:CovEffectOM}
\begin{tabular}{ll S[table-format=-1.4] S[table-format=-1.4] S[table-format=-1.4] }
\toprule
\textsc{covariate}   & &  \text{$\Delta$P(not legal)} &  \text{$\Delta$P(medicinal use)}
     &  \text{$\Delta$P(personal use)}   \\
\midrule
\textsc{age, 10 years}       & &   0.015   &    0.012   & -0.028      \\
\textsc{income, \$10,000}    & &  -0.005   &   -0.003   &  0.008      \\
\textsc{past use}            & &  -0.129   &   -0.113   &  0.243      \\
\textsc{bachelors \& above}  & &  -0.045   &   -0.035   &  0.080      \\
\textsc{eventually legal}    & &  -0.126   &   -0.060   &  0.186      \\
\textsc{other races}         & &   0.059   &    0.031   & -0.089      \\
\textsc{democrat}            & &  -0.080   &   -0.066   &  0.147      \\
\textsc{other parties}       & &  -0.070   &   -0.051   &  0.121      \\
\textsc{liberal}             & &  -0.068   &   -0.066   &  0.134      \\
\bottomrule
\end{tabular}
\end{table}

We mentioned in Section~\ref{sec:Model1} that the coefficients of the ordinal
probit model only give the direction of impact for the first and last
categories, but not the remaining categories. The actual covariate effects
need to be calculated for all the categories. We compute the average
covariate effects for all significant variables and present them in
Table~\ref{Table:CovEffectOM}. From the table, note that past use,
eventually legal, and identifying oneself as a Democrat are three variables
with the highest impact on public opinion. Past use of marijuana increases
support for personal use by 24.3 percent and decreases the support for
medicinal use and oppose legalization by 11.3 and 12.9 percents,
respectively. Similarly, a respondent who expects marijuana to be legal is
18.6 percent more likely to support marijuana for personal use. This increase
comes from a decrease in probability for medicinal use and oppose
legalization, which are 6.0 and 12.6 percents, respectively. In the same way,
a respondent who is a Democrat is 14.7 percent more likely to favor personal
use, and 6.6 and 8.0 percents less likely to favor medicinal use and oppose
legalization, respectively. The covariate effects for the remaining variables
can be interpreted similarly.

\section{Conclusion}\label{sec:Conclusion}

This chapter presents an overview of two popular ordinal models (ordinal
probit and ordinal logit models) as well as two widely used binary models
(probit and logit models). These models fall within the class of discrete
choice models and are extremely popular in several disciplines including
economics, epidemiology, finance, and sociology. The models are described
using the latent variable threshold-crossing framework since it elegantly
connects individual choice behavior with the random utility model in
economics. While we focus on the ordinal probit and binary probit models as
prototypes to derive the likelihood and outline the estimation procedure, the
approach is completely applicable to its logit counterparts. Some interesting
aspects about interpreting the coefficients of logit models are emphasized.
Since the models considered here are non-linear models, the coefficients
cannot be interpreted as covariate effects. We explain how to compute the
covariate effects when the covariates are continuous and when they are binary
(indicator variable). Measures to assess model fitting, namely, likelihood
ratio statistic, McFadden's R-square, and hit-rate are also described. We
also include specific applications of discrete choice models, wherein we
utilize the ordinal probit and binary probit models to analyze public opinion
on marijuana legalization and extent of legalization in the United States --
a notable and contentious topic with credible arguments from proponents as
well as critics of the policy.

The first application utilizes a binary probit model to analyze the response
to legalization (`oppose legalization' or `favor legalization') based on
individual demographic variables, educational background, racial
characteristics and political affiliation of the respondents. The data is
taken from the March 2013 Political Survey collected by the Pew Research
Center. The results suggest that while log age has a negative effect, past
use of marijuana, male, higher education, affiliation to the Democratic or
`Independent \& Other' parties, as well as Roman Catholic, Conservative and
Liberal religious beliefs have a positive effect on the probability of
supporting legalization. Not surprisingly, past use of marijuana has the
highest positive effect and increases the probability of support by 28.5
percent. The proposed model performs well and correctly classifies 69 percent
of the responses.

The second study employs an ordinal probit model to analyze the ordered
response to legalization (`oppose legalization', `legal for medicinal use' or
`legal for personal use') based on a similar set of covariates as in the
first study. Data for this study is taken from the February 2014 Political
Survey collected by the Pew Research Center. The results show that log age
and belonging to `Other Races' (non-White and non-African American)
negatively (positively) affects the probability of supporting personal use
(oppose legalization) and the latter has the highest negative effect at $8.9$
percent. The variables that have a positive (negative) effect on personal use
(oppose legalization) include income and indicators for past use of
marijuana, bachelors or higher education, individual expectation on eventual
legalization, Democratic Party, `Other Parties' and Liberal religious
beliefs. Amongst these, past use of marijuana has the highest positive effect
on personal use at 24.3 percent. The proposed ordinal model performs well and
correctly classifies approximately 59 percent of the responses.

The insights from these studies are interesting and may assist policymakers
to better assess public preference regarding marijuana legalization, the
extent of legalization (particularly, medical marijuana) and the factors
associated with such preferences. For instance, the finding that educational
attainment has a positive impact on public support for legalization implies
that providing information on the medicinal findings on marijuana and
emphasizing college and university education is likely to increase support
for it. If people believe that the benefits outweigh the costs, then public
opinion will move further towards legalization and vice-versa. These findings
may also be helpful to various advocacy groups and opposition lobbies engaged
in promoting or opposing unrestricted legalization of marijuana respectively.
A clear understanding of the underlying factors that drive an individual's
opinion will help these groups better plan their campaigns. For example,
since support for legalization is negatively related with age, groups
opposing legalization may consider not campaigning amongst the youth as it is
unlikely to yield support. Similarly, lobbies engaged in opposing
legalization may optimize their time use on better possibilities than
convince an individual or group with a history of marijuana use to oppose
legalization.

\clearpage \pagebreak
\pdfbookmark[1]{References}{unnumbered}    

\bibliography{MarijuanaBib}
\bibliographystyle{jasa}



\end{document}